\documentstyle[aps,12pt]{revtex}
\begin{document}
\title{Metallic charge density waves and surface Mott insulators 
for adlayer structures on semiconductors: extended Hubbard modeling} 
\vspace{10mm}
\author{Giuseppe Santoro$^{1,2}$, Sandro Sorella$^{1,2}$, 
Federico Becca$^{1,2}$, Sandro Scandolo$^{2,3}$, and 
Erio Tosatti$^{1,2,3}$}
\address{
$^{(1)}$ International School for Advanced Studies (SISSA), Via Beirut 2,
Trieste, Italy\\
$^{(2)}$ Istituto Nazionale per la Fisica della Materia (INFM), 
Via Beirut 2, Trieste, Italy\\
$^{(3)}$ International Center for Theoretical Physics (ICTP), Strada Costiera,
Trieste, Italy}
\maketitle
\vspace{10mm}
\begin{abstract}
Motivated by the recent experimental evidence of commensurate surface CDW 
in Pb/Ge(111) and Sn/Ge(111) $\surd{3}$-adlayer structures, 
as well as by the insulating states found on K/Si(111):B and SiC(0001),
we have investigated the role of electron-electron interactions, and also of 
electron-phonon coupling, on the narrow surface state band originating
from the dangling bond orbitals of the adsorbate.
We model the problem by an extended two-dimensional Hubbard model at
half-filling on a triangular lattice. We include an on-site Hubbard
repulsion $U$ and a nearest-neighbor $V$, plus a long-ranged Coulomb tail.
The electron-phonon interaction is treated in the deformation potential
approximation.
We have explored the phase diagram of the model including the possibility 
of commensurate $3\times 3$ phases, using mainly the Hartree-Fock 
approximation. 
For $U$ larger than the bandwidth we find magnetic insulators,
possibly corresponding to the situation in SiC and in K/Si.
For smaller $U$, the inter-site repulsion $V$ can stabilize metallic
CDW phases, reminiscent of the $3\times 3$ structures of Sn/Ge, and
possibly of Pb/Ge. 
\end{abstract}
\newpage

\section{Introduction}

Pb and Sn $\sqrt{3}$-adlayer structures on the (111) surface of Ge have
recently revealed a charge density wave (CDW) instability to a low temperature 
reconstructed $3\times 3$ phase.\cite{nature,Modesti,Avila,LeLay,Sn_carpinelli} 
The low temperature phase is either metallic -- as seems to be the case
for Sn/Ge(111) -- or weakly gapped, or pseudo-gapped, as suggested
for Pb/Ge(111). Related systems, like the $\sqrt{3}$-adlayer of Si on
the (0001) surface of SiC \cite{SiC} and on K/Si(111):B,\cite{KSi}
show, instead, a clear insulating behavior -- with a large gap --
with no structural anomalies or CDWs investigated so far. 
Two-dimensional Fermi surface (FS) nesting in the half-filled surface 
states\cite{tosatti} has been invoked as the driving mechanism for the
CDW instability in the case of Pb/Ge,\cite{nature} but excluded for the 
case of Sn/Ge.\cite{Sn_carpinelli}

As all these systems appear to belong together in the same class, we should
begin with a discussion that in principle encompasses all of them.
We believe the following points to be of general validity:
{\em i)\/} {\bf Poor nesting}. An unbiased interpretation of the photoemission
experiments,\cite{Modesti,Avila,LeLay} and a close examination of the existing 
local density approximation (LDA) calculations\cite{nature,Sandro,Sn_carpinelli}
of the surface (``adatom dangling bond'') half-filled band do not indicate
good nesting of the FS at the surface Brillouin zone (BZ) corner 
${\bf K}=(4\pi/3a,0)$. 
We believe this to be equally true for Pb/Ge as for Sn/Ge, contrary to 
what stated in the literature.\cite{nature} 
Fig.\ 1, showing the LDA surface band dispersion 
for the test-case of Si(111)/Si,\cite{Sandro} as well as the
corresponding FS and Lindhard density response function,
provides a concrete illustration of these statements.
{\em ii)\/} {\bf Importance of electron-electron interactions}. 
The width $W$ of the surface band is relatively small 
($W\approx 0.5$ eV for Pb-Sn/Ge, $\approx 0.3$ eV for SiC). 
Moreover, this band is half-filled.
These facts call for a careful consideration of electron-electron
interactions, and not just electron-phonon (e-ph), as a possible source of 
instability. This is reinforced by noting the different phenomenology
of SiC and K/Si:B with respect to Pb-Sn/Ge, the stronger insulating
character of the former paralleling closely their stronger electron-electron 
repulsions, connected with larger bulk semiconducting gaps.
{\em iii)\/} {\bf Weakness of LDA}. LDA is certainly suitable for calculating 
realistic bands in a weakly interacting system,
but is less reliable, at least without including spin, in predicting the 
instabilities of a narrow band. 
For instance, the phenomenology of SiC(0001) 
-- suggesting a Mott-Hubbard insulator -- is unreproducible by LDA. 
The very onset of a CDW on Sn/Ge(111) does not seem to be predicted by recent 
LDA calculations.\cite{Sn_carpinelli}
{\em iv)\/} {\bf Different mechanisms for $3\times 3$ CDW instabilities}.
There are at least two different mechanisms which can influence the CDW
formation:
a) on-site, and nearest-neighbor (n.n.) electron-electron repulsion
b) on-site effective attraction (negative Hubbard-$U$ term) of
e-ph origin.
Of these, the n.n. repulsion naturally suggests, as we shall see, 
the $3\times 3$ surface periodicity, which is found experimentally. 
Electron-phonon alone would not in itself appear to drive a $3\times 3$ CDW.
At weak coupling, the band structure dominates and incommensurate
structures could be preferred. At strong coupling, the frustration
associated to the triangular lattice, will favor, in general, a
superconducting ground state over a CDW phase.\cite{santos}.

The approach we take here is based on an extended Hubbard-Holstein model.
It is by necessity a ``non-first-principle'' approach and, as such,
it has no strong predictive power. We have found it very helpful, however,
in clarifying the possible scenarios as a function of physical parameters.

\section{Model}

The basic starting point is the half-filled surface state band 
$\epsilon_{\bf k}$ which one calculates in LDA, and is found to lie in the
bulk gap.\cite{nature,Sandro}
We write our effective Hamiltonian as follows:
\begin{equation}
H \,=\, \sum_{{\bf k}}^{BZ} \sum_{\sigma} \epsilon_{\bf k} 
c^{\dagger}_{{\bf k},\sigma} c_{{\bf k},\sigma} \,+\, H_{\rm ph} \,+\, 
H_{\rm e-ph} \,+\, H_{\rm int} \;,
\end{equation}
where $c^{\dagger}_{{\bf k},\sigma}$ is the Bloch surface state, 
whose Wannier function is centered on the adatom, as can be seen in 
Ref.\ \cite{Sandro}.
$H_{\rm int}$ includes correlation effects which are not correctly accounted 
for within LDA, which we parametrize as follows:
\begin{equation}
H_{\rm int} = U \sum_{{\bf r}} n_{{\bf r},\uparrow} n_{{\bf r},\downarrow}
+ \frac{1}{2} \sum_{{\bf r}\ne {\bf r}'} V_{{\bf r}-{\bf r}'} 
(n_{{\bf r}}-1) (n_{{\bf r}'}-1) \;.
\end{equation}
Here $U$ is an effective repulsion (Hubbard-$U$) for two electrons on the same 
adatom Wannier orbital, and $V_{{\bf r}-{\bf r}'}$ is the 
direct Coulomb interaction between different sites ${\bf r}$ and ${\bf r}'$.
We have considered two models for $V_{{\bf r}-{\bf r}'}$: 
1) a truncation to a n.n. $V$, and 2) a model with a Coulomb tail 
$V_{{\bf r}-{\bf r}'}=Va/{|{\bf r}-{\bf r}'|}$. 
The results for case 2) are qualitatively similar to those of 1), 
and will be discussed elsewhere.\cite{santoro}
LDA estimates of the bare repulsions $U_o$ and $V_o$ between two electrons on 
the same and on neighboring Wannier orbitals are 
-- for our test case of Si(111)/Si -- of about $3.6$ eV and 
$1.8$ eV respectively.\cite{Sandro} 
Screening effects by the the underlying bulk are expected to
reduce very substantially these repulsive energies. A conservative lower
bound for $U$ and $V$ is obtained dividing their bare values
by the image-charge screening factor, $(\epsilon +1)/2\approx 6$, 
yielding $U=0.6$ eV, and $V=0.3$ eV.
As for the e-ph interaction, in principle both the on-site
Wannier state energy and the hopping matrix elements between first neighbors
depend on the positions of the adatoms. 
Within the deformation potential approximation, we consider only a 
linear dependence of the on-site energy from a single ionic coordinate 
(for instance, the height $z_{\bf r}$ of the adatom measured 
from the equilibrium position), and take 
$H_{\rm e-ph}=-g\sum_{{\bf r}} z_{\bf r} (n_{\bf r}-1)$.
The free-phonon term will have the usual form
$H_{\rm ph} = \sum_{\bf k} \hbar \omega_{\bf k} 
\left( b^{\dagger}_{\bf k} b_{\bf k} + 1/2 \right)$, 
where $b_{\bf k}$ is the phonon annihilation operator. 
An order-of-magnitude estimate for $g$ is $\approx 1$ eV/$\AA$. 

\section{Phase diagram.} 
\label{pd:sec}

We consider first the purely electronic problem in the absence of 
e-ph interaction. 
We start the discussion from particular limiting cases for which 
exact statements, or at least well-controlled ones, can be made.

{\bf Large positive $U$: the Mott insulator.}
For $U\gg V,W$, the system is deep inside the Mott insulating 
regime.\cite{Anderson_SE}
Within the large manifold of spin degenerate states with exactly one electron 
per site, the kinetic energy generates, in second order perturbation theory, 
the Heisenberg spin-1/2 antiferromagnet as the effective Hamiltonian
governing the {\it spin\/} degrees of freedom,
$H_{\rm eff}=\sum_{(ij)} J_{ij} {\bf S}_{{\bf r}_i} \cdot {\bf S}_{{\bf r}_j}$, 
with $J_{ij}=4|t_{ij}|^2/U$.\cite{Anderson_SE}
For our test case of Si(111)/Si, the values of the hoppings are such that 
$J_2/J_1\approx 0.12$ while the remaining couplings $J_3,\cdots$ are very small.  
Antiferromagnetism is frustrated on the triangular lattice. 
Zero temperature long range order (LRO) of the three-sublattice $120^o$-N\'eel 
type -- a commensurate spiral spin density wave (s-SDW) --
is nevertheless likely to occur for the spin 1/2 antiferromagnetic
Heisenberg model with n.n. coupling $J_1$, and for small enough $J_2$.
In summary, we expect for large values of $U$ a wide-gap Mott insulator with 
a s-SDW (spins lying in a plane, forming $120^o$ angles), 
and a $3\times 3$ {\it magnetic} unit cell. 
This is, most likely, the state to be found on the Si-terminated SiC(0001) 
surface at T=0.

{\bf Strong inter-site repulsion:
an asymmetric CDW with three inequivalent sites.}
The e-ph coupling can effectively reduce $U$, but not $V$. 
Therefore, it is of interest to consider the hypothetical regime
$W<U\ll V$.
In order to minimize the interaction energy, the system will prefer a
$3\times 3$ CDW with two electrons on one sublattice (A), 
a single electron on another sublattice (B), and
zero electrons on the third sublattice (C). (See Fig.\ 2.)
The spin degeneracy associated with the (unpaired) electron
on sublattice B can be removed, owing to $t_2$, which leads to an 
effective spin-1/2 Heisenberg Hamiltonian within sublattice B, 
with a weak antiferromagnetic exchange constant $J=4t^2_2/U$.\cite{santoro}.
Summarizing, we expect in this regime a strong $3\times 3$ asymmetric CDW 
with three inequivalent sites (a-CDW), and a spiral $3\sqrt{3}\times 3\sqrt{3}$ SDW, 
governing the unpaired electron spins, superimposed on it.  
This a-CDW is not compatible with the experimental findings on Pb-Sn/Ge,
but it could be realized in some other case.

{\bf Mean-field theory}. 
In order to get an idea of any additional phases for smaller 
$U$, and of the possible phase diagram of the model we turn to a mean field 
analysis.
The first issue is the possibility of magnetism.
For small values of the interactions $U$ and $V$, 
the Stoner criterion can be used to 
study the instability of the paramagnetic metal obtained from LDA calculations.
The charge and spin susceptibility are given, within the random phase
approximation, by 
$\chi_C({\bf q}) = 2\chi_o({\bf q})/[1 + (U+2V_{\bf q})\chi_o({\bf q})]$, and
$\chi_S({\bf q}) = \chi_o({\bf q})/[1 - U\chi_o({\bf q})]$, 
where $\chi_o$ is the non-interacting susceptibility. 
The divergence of $\chi_S$ is governed, in this approximation, by $U$ only.
Since $\chi_o({\bf q})$ is finite everywhere, a finite $U$ is needed in
order to destabilize the paramagnetic metal. 
The wavevector ${\bf q}^*$ at which $\chi_S$ first diverges, by increasing $U$,
is in general incommensurate with the underlying unit cell.
The instability is towards an incommensurate metallic spiral SDW.\cite{KMurthy}
As for the charge susceptibility, a divergence can be caused only by
an attractive Fourier component of the potential $V_{\bf q}$. 
$V_{\bf q}$ has a minimum at the BZ corners $\pm {\bf K}$, with
$V_{\pm {\bf K}}=-3V$. 
($V_{\pm {\bf K}}\approx -1.54 V$ if a Coulomb tail is added). 
This minimum leads to an instability towards a $3\times 3$ CDW as 
$(U+2V_{{\bf K}})\chi_o({\bf K})=-1$. 

In general, the small coupling paramagnetic metal is surrounded by an 
intermediate coupling region, where complicated incommensurate --
generally metallic -- solutions occur. 
For stronger $U$ and $V$, commensurate solutions are privileged.\cite{KMurthy}
In view of the fact that a $3\times 3$ CDW is experimentally relevant, we
concentrate our analysis on the simplest commensurate phases. 
These are easy to study with a standard Hartree-Fock (HF) mean-field 
theory.\cite{santoro}
In particular, we restrict ourselves to order parameters
associated with momentum space averages of the type
$\langle c^{\dagger}_{{\bf k},\sigma} c_{{\bf k},\sigma '} \rangle$ and 
$\langle c^{\dagger}_{{\bf k},\sigma} c_{{\bf k}\pm{\bf K},\sigma '}\rangle$,  
i.e., the uniform magnetization density ${\bf m}$, the ${\bf K}$-component of 
the charge density 
$\rho_{\bf K} = (1/N) \sum_{{\bf k},\sigma} 
\langle c^{\dagger}_{{\bf k},\sigma} c_{{\bf k}-{\bf K},\sigma} \rangle$,
and the ${\bf K}$-component of the spin density
${\bf S}_{\bf K} = (1/2N) \sum_{{\bf k}} \sum_{\alpha,\beta}
\langle c^{\dagger}_{{\bf k},\alpha} ({\vec \sigma})_{\alpha\beta} 
c_{{\bf k}-{\bf K},\beta} \rangle$.
$\rho_{\bf K}$ and ${\bf S}_{\bf K}$ have phase freedom, and 
are generally complex: $\rho_{\bf K}=|\rho_{\bf K}| e^{i\phi_{\rho}}$, etc.
The role of the phase is clarified by looking at the real-space 
distribution within the $3\times 3$ unit cell. 
For the charge, for instance, 
$\langle n_{{\bf r}_j}\rangle=1+2|\rho_{\bf K}|\cos{(2\pi p_j/3+\phi_{\rho})}$,
where $p_j=0,1,2$, respectively, on sublattice A, B, and C. 
The e-ph coupling is included but, after linearization, the
displacement order parameter is not independent, and is given by
$\langle z_{\bf K}\rangle=(g/M\omega_{\bf K}^2)\rho_{\bf K}$.
Only the phonon modes at $\pm{\bf K}$ couple directly to the CDW. 
The phonon part of the Hamiltonian can be diagonalized by displacing
the oscillators at $\pm{\bf K}$. 
This gives just an extra term in the electronic HF Hamiltonian of the form
$\Delta U (\rho^*_{\bf K} {\hat \rho}_{\bf K} + {\rm H.c.})$, 
with an energy $\Delta U=-g^2/M\omega^2_{\bf K}$ which is the relevant 
coupling parameter.  
This term acts, effectively, as a negative-$U$ contribution acting only
on the charge part of the electronic Hamiltonian.

The mean-field solutions must be compatible with the symmetry of the
problem.
A symmetry analysis of the Landau theory \cite{Toledano} built from 
the order parameters ${\bf m}$, $\rho_{\bf K}$, and ${\bf S}_{\bf K}$ 
shows that:\cite{santoro}
{\it i\/}) A CDW can occur without concomitant magnetism. 
{\it ii\/}) A SDW can occur without a CDW only in the form of a $120^o$
spiral SDW. In all other cases, a SDW {\em implies\/} also a CDW.
{\it iii\/}) The simultaneous presence of a SDW and a CDW implies, generally, 
a finite magnetization $\bf m$, unless the phases of 
$\rho$ and $S$ are such that $2\phi_{\rho}+\phi_{\sigma}=\pi/2+n\pi$. 

We present a brief summary of the mean-field HF calculations for 
arbitrary $U$, $V$, and $g$.
The main phases present in the HF phase diagram are shown in Fig.\ 2 
for the case of $g=0$. (The diagram is qualitatively similar for $g\ne 0$.)
The s-SDW dominates the large $U$, small $V$ part of the phase diagram, 
as expected from the Heisenberg model mapping at $U\to\infty$. 
This is the Mott insulator phase, probably relevant for SiC.
Its HF bands are shown in Fig.\ 3(a).
There is however another solution of the HF equations in the large $U$, 
small $V$ region. 
It is an insulating state characterized by a linear z-SDW plus a small 
CDW with $\phi_{\rho}=0$, accompanied by a magnetization $m^z=1/3$. 
It lies above the s-SDW by only a small energy difference 
(of order $t_1$ per site), and it could be stabilized by other factors
(e.g., spin-orbit). 
A recent LSDA calculation for $\sqrt{3}$-Si/Si(111) has indicated this
z-SDW as the ground state, at least if spins are forced to be 
collinear.\cite{Sandro}
The HF bands for this solution are shown in Fig.\ 3(b). 
By increasing $V$, the energies of the s-SDW and of the z-SDW tend to approach, 
until they cross at a critical value $V_c$ of $V$. 
We find $V_c/t_1\approx 3.3$ at $U/t_1=10$.
As $V>V_c$, however, the insulating a-CDW prevails. 
For small values of $U$ and $V$, or for large enough e-ph coupling
$g$, a {\em metallic\/} CDW with $\phi_{\rho}=0$ (m-CDW) is found. 
This may be relevant for the case of Pb and of Sn/Ge(111). 
(See Fig.\ 3(c) for the HF bands.)
The degree of metallicity of this phase is much reduced relative to the 
undistorted surface (pseudo-gap). 
We note that the e-ph interaction can stabilize the 
$\phi_{\rho}=0$ m-CDW also at relatively large $U$, by countering $U$ with a
large negative $\Delta U$.
With $g=1$ eV/$\AA$, $M_{Si}=28$, and $\omega_{\bf K}\approx 30$ meV
we get $\Delta U \approx -3 t_1$, sufficient to switch from a 
s-SDW ground state to a m-CDW for, say, $U/t_1=8$ and $V/t_1=2$. 
Much larger $g$'s would eventually stabilize a superconducting 
ground state.\cite{santoro}

In conclusion, we have found within a single phase diagram three phases,
the s-SDW, the z-SDW, and the m-CDW, which may be relevant, respectively,
to SiC(0001), to K/Si(111):B, and to Pb or Sn/Ge(111).  

We acknowledge financial support from INFM, through projects LOTUS and
HTSC, and from EU, through ERBCHRXCT940438.
We thank S. Modesti, M.C. Asensio, J. Avila, G. Le Lay, and E.W. Plummer
for discussions.


\newpage
\begin{center} {\bf Figure Captions} \end{center}

\begin{description}

\item[Figure 1]
Surface state dispersion for Si(111)/Si, as obtained from LDA 
(solid squares). The solid line is a tight-binding fit obtained by
including up to the sixth shell of neighbors, $t_1,\cdots,t_6$. 
The fit gives $t_1=0.0551$ eV, and $t_2/t_1=-0.3494$, $t_3/t_1=0.1335$, 
$t_4/t_1=-0.0615$, $t_5/t_1=0.0042$, $t_6/t_1=-0.0215$. 
Upper inset: The Fermi surface of the half-filled surface band. 
Notice the quite poor nesting at the BZ corner wavevector ${\bf K}=(4\pi/3a,0)$. 
Lower inset: The zero temperature Lindhard function $\chi_o(\bf q)$
for the half-filled surface band. 
Notice the two peaks located at ${\bf q}_1\approx 0.525 {\bf K}$ and 
${\bf q}_2\approx 1.32 {\bf K}$, and no feature whatsoever at ${\bf K}$. 

\item[Figure 2]
Schematic Hartree-Fock phase diagram of the $U-V$ model, at zero
electron-phonon coupling, for the band structure shown in Fig.\ 1.
Only the most important commensurate $3\times 3$ phases have been studied. 

\item[Figure 3]
Plot of the HF bands along high symmetry directions of the BZ
for the s-SDW and two CDW $\phi_{\rho}=0$ solutions: 
(a) at $U/t_1=9$ and $V/t_1=2$, the insulating s-SDW (ground state);
(b) at $U/t_1=9$ and $V/t_1=2$, the insulating solution with a small
CDW and $m^z=1/3$ (meta-stable, the actual ground state being the s-SDW);
Solid and dashed lines denote up and down bands, respectively.
(c) at $U/t_1=4$ and $V/t_1=2$, the metallic solution with a large
CDW and no magnetism. 

\end{description}


\begin{thebibliography}{99}

\bibitem{nature}
J. M. Carpinelli {\em et al.\/}, Nature {\bf 381}, 398 (1996).

\bibitem{Modesti}
A. Goldoni, C. Cepek, S. Modesti, Phys.\ Rev.\ B {\bf 55}, 4109 (1997); 
S. Modesti (private commun.).

\bibitem{Sn_carpinelli}
J. M. Carpinelli, H. H. Weitering, M. Bartkowiak, R. Stumpf, and 
E. W. Plummer (preprint).

\bibitem{Avila}
J. Avila, A. Mascaraque, E.G. Michel, and M.C. Asensio (to be published).

\bibitem{LeLay}
G. Le Lay {\em et al.\/} (to be published). 

\bibitem{SiC}
L. I. Johansson {\em et al.\/}, Surf.\ Sci.\ {\bf 360}, L478 (1996);
J.-M. Themlin {\em et al.\/}, Europhys.\ Lett.\ {\bf 39}, 61 (1997).

\bibitem{KSi}
H. H. Weitering {\em et al.\/}, Phys.\ Rev.\ Lett.\ {\bf 78}, 1331 (1997). 

\bibitem{tosatti}
E. Tosatti and P. W. Anderson, in Proc.\ 2nd Int.\ Conf.\
on Solid Surfaces, ed. S. Kawaji, Jap. J. Appl.\ Phys., Pt.\ 2, Suppl.\ 2,
381 (1974);
E. Tosatti, in {\em Electronic surface and interface states on metallic 
systems\/}, p.\ 67, eds. E. Bertel and M. Donath (World Scientific, 
Singapore, 1995). 

\bibitem{Sandro}
S. Scandolo {\em et al.\/}, proceedings of ECOSS-17.

\bibitem{santos}
R. R. dos Santos, Phys.\ Rev.\ B {\bf 48}, 3976 (1993). 

\bibitem{santoro}
G. Santoro {\em et al.\/} (in preparation).

\bibitem{Anderson_SE}
P. W. Anderson in {\em Frontiers and Borderlines in Many-Particle Physics\/},
Proc.\ E. Fermi Summer School in Varenna, july 1987 (North-Holland,
Amsterdam, 1988).

\bibitem{KMurthy}
H. R. Krishnamurthy {\em et al.\/}, Phys.\ Rev.\ Lett.\ {\bf 64}, 950 (1990); 
C. Jayaprakash {\em et al.\/}, Europhys.\ Lett.\ {\bf 15}, 625 (1991). 

\bibitem{Toledano}
See J. Tol\'edano and P. Tol\'edano, 
{\em The Landau theory of phase transitions\/},
(World Scientific, Singapore, 1987). 

\end{thebibliography}
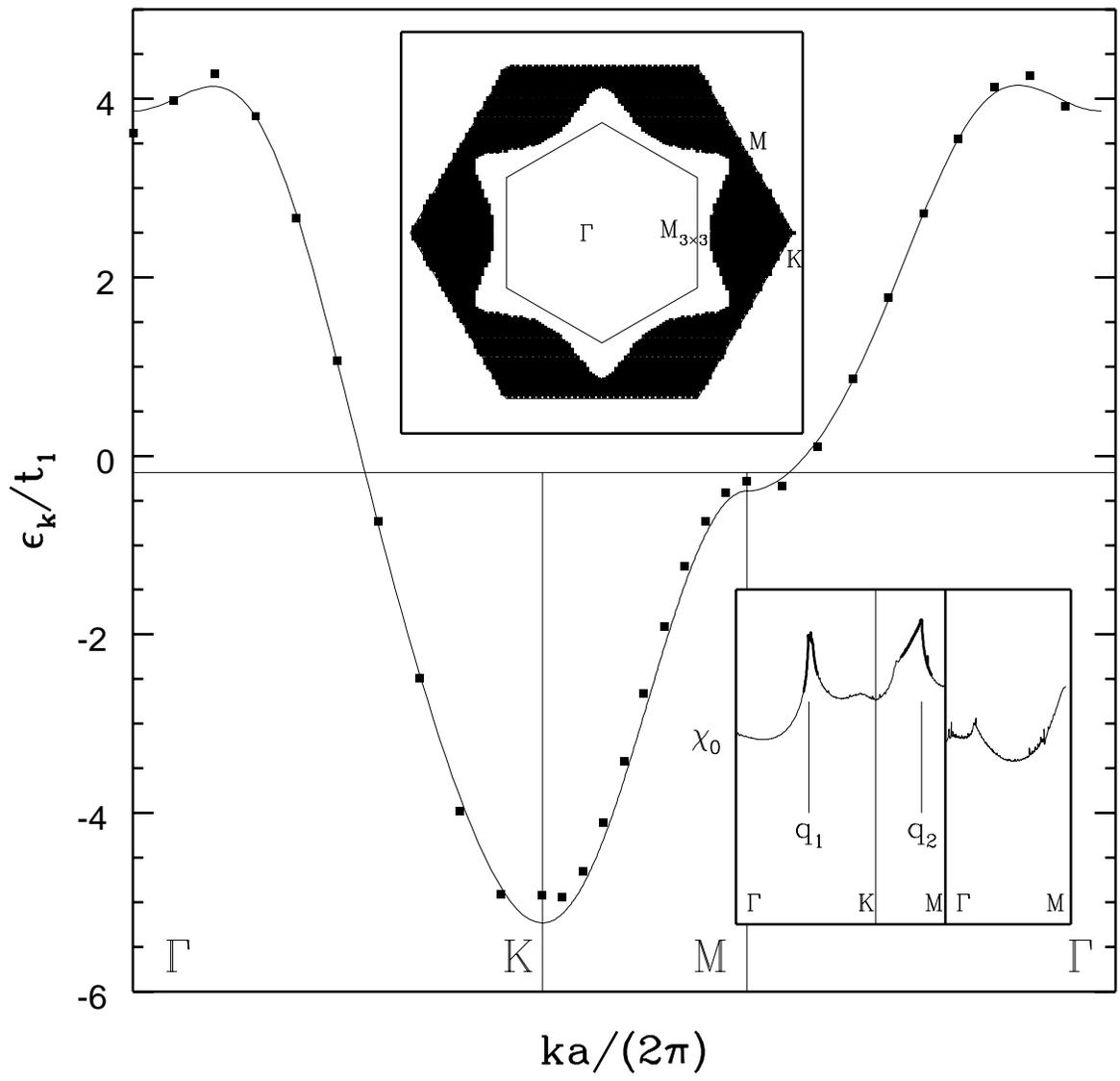
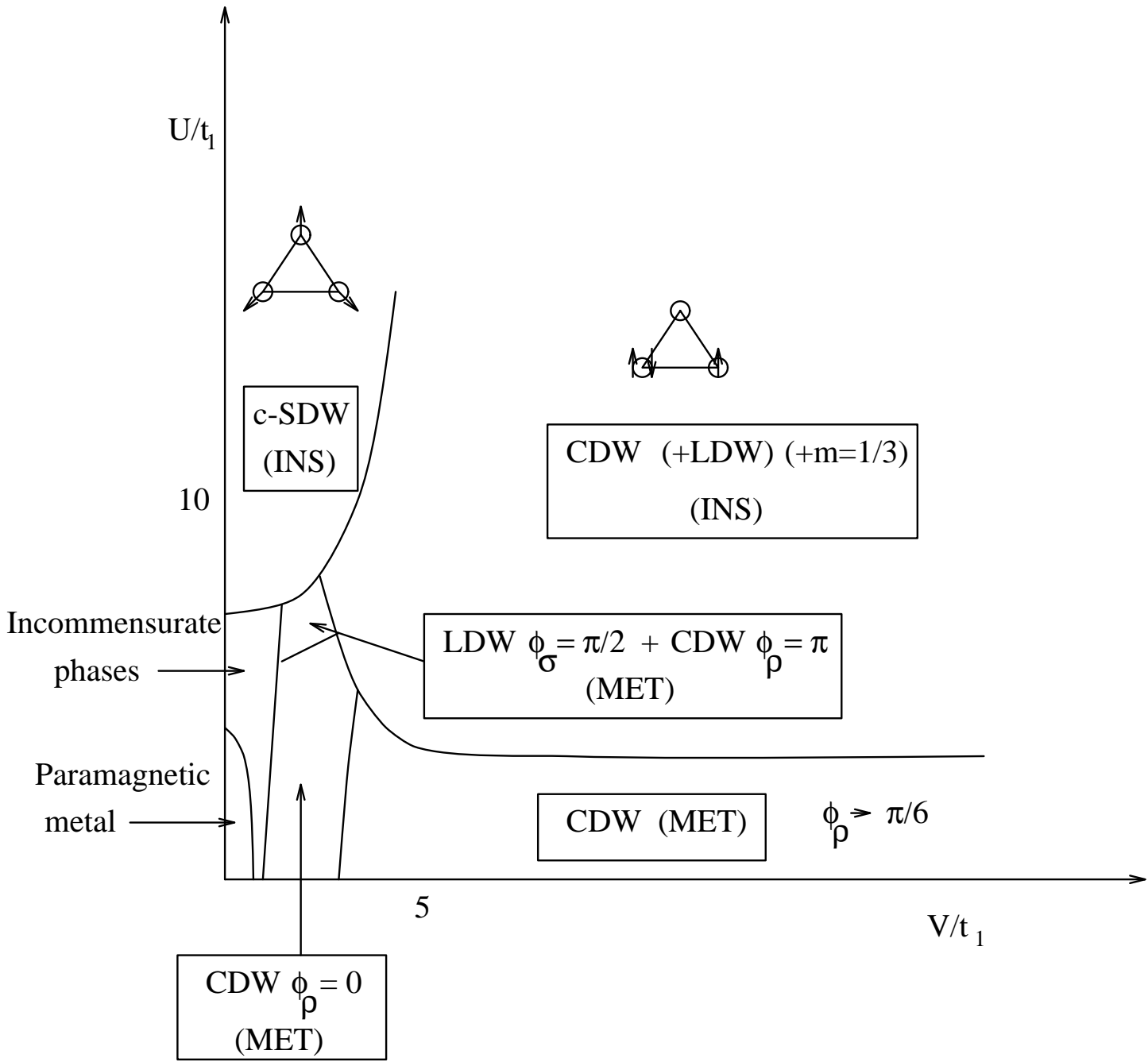
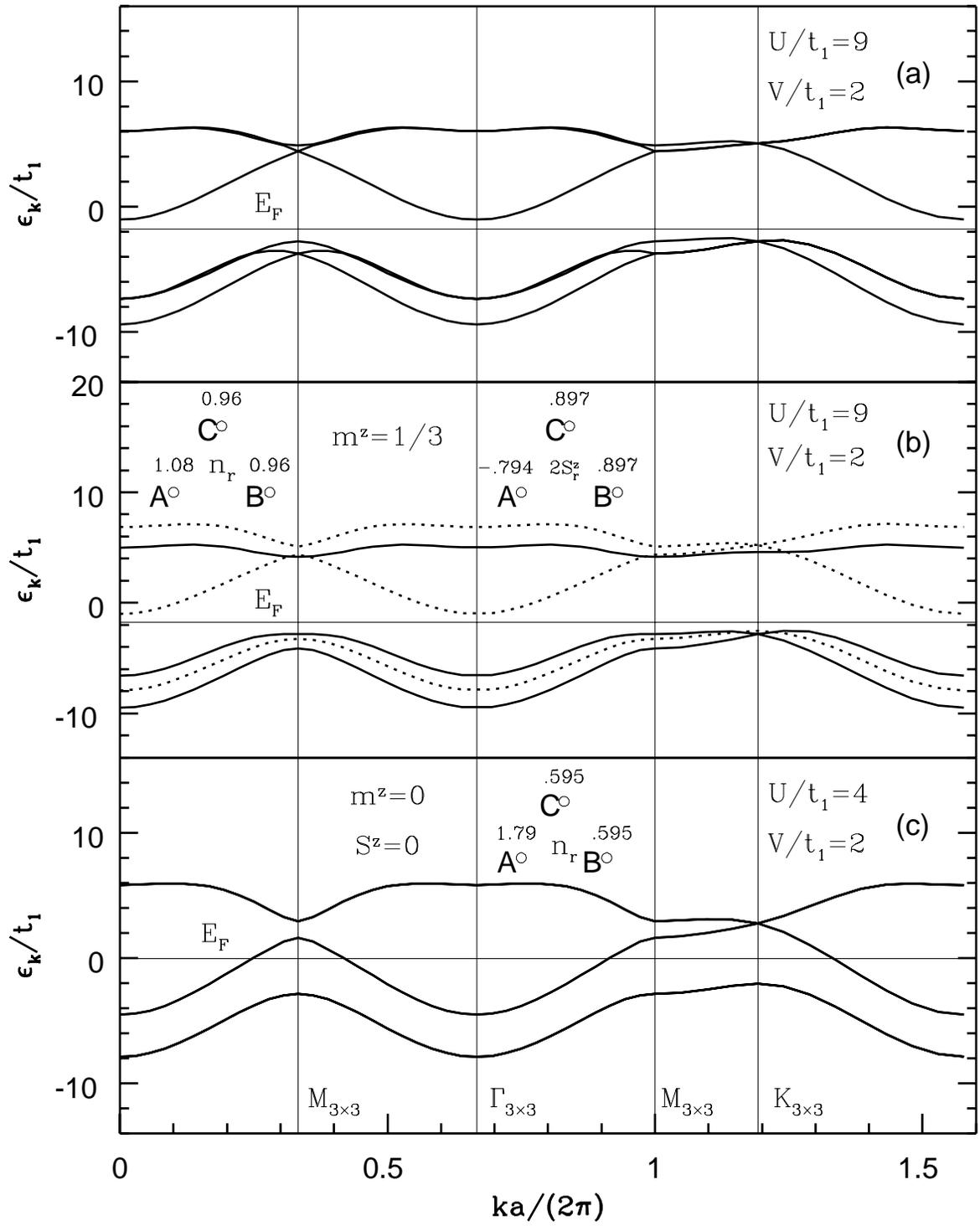
\end{document}